\documentclass[conference]{IEEEtran}

\usepackage{graphicx}
\usepackage{amsmath}
\usepackage{balance}
\usepackage{cite}
\usepackage{url}
\usepackage[inline]{enumitem} % For inline enumerations

\usepackage{vmr-symbols-vecbold}
\usepackage{standard-macros}
\PassOptionsToPackage{hyphens}{url}
\usepackage{xcolor}
\definecolor{links}{rgb}{0,0,0.9}   % blue
\definecolor{urls}{rgb}{0,0,0.9}    % blue
\definecolor{cites}{rgb}{0,0,0.9}   % blue
\usepackage[colorlinks,hyperindex,linkcolor=links,citecolor=cites,urlcolor=urls]{hyperref} % generates colored links in pdf file

\DeclareSymbolFontAlphabet{\amsmathbb}{AMSb}%

\newcommand{\lro}[1]{\lefto({#1}\right)}																% left right paranthesis operator
\newcommand{\lr}[1]{\left({#1}\right)}																% left right paranthesis operator
\newcommand{\lrb}[1]{\left \lbrace {#1} \right \rbrace}															% left right braces operator
\safemath{\dopplerspread}{B_D}																								% doppler spread
\safemath{\delayspread}{T_D}																									% delay spread
\safemath{\nc}{n\sub{c}}																										% coherence time
\safemath{\nf}{n\sub{f}}																										% feedback message length
\safemath{\efa}{p\sub{sc}}
\safemath{\efb}{p\sub{cs}}
\safemath{\ef}{\epsilon\sub{f}	}
\safemath{\nd}{n\sub{d}}																										% data symbols
\safemath{\ntx}{n\sub{t}} 																											% transmit antennas
\safemath{\nrx}{n\sub{r}}																											% receive antennas
\safemath{\ntxt}{\tilde{n\sub{t}}}																											% receive antennas
\safemath{\cb}{\ensuremath{L}} 																								% code blocks
\safemath{\cl}{\ensuremath{n}} 																								% codelength
\safemath{\txanto}{{\ensuremath{\tilde{m}_t}}} 																		% transmit antennas when some is turned off
\safemath{\cs}{M} 																														% code size
\safemath{\idPustm}{\ensuremath{S_{k}}}
\safemath{\error}{\ensuremath{\epsilon}} 																				%Error target
\safemath{\eexp}{\ensuremath{\mathcal{E}}} 																			%Error exponent
\safemath{\nsubc}{n\sub{s}}			 																						% number of subcarriers
\safemath{\nofdm}{n\sub{o}} 																									% number of OFDM symbols
\safemath{\bc}{\ensuremath{B_c}} 																							% Coherence bandwidth
\safemath{\ts}{\ensuremath{T_s}} 																							% Symbol time
\safemath{\nrb}{\ensuremath{n_{rb}}} 																						% Symbol time
\safemath{\rul}{\ensuremath{\rho\sub{ul}}}
\safemath{\rdl}{\ensuremath{\rho\sub{dl}}}

\safemath{\nres}{\ell}
\safemath{\nr}{n\sub{r}}
\safemath{\maxk}{M^*\lr{\nres, \nsubc, \nofdm, \epsilon, \rho}}
\safemath{\Rmax}{R^*}%\lr{\nres, \nsubc, \nofdm,M, \epsilon, \rho}}
\safemath{\Emin}{E\sub{b}^*/N_0}%\lr{\nres, \nsubc, \nofdm,M, \epsilon, \rho}}
\safemath{\Eminf}{\frac{E\sub{b}^*}{N_0}}
\safemath{\np}{\ensuremath{n\sub{p}}}
\safemath{\ndf}{\ensuremath{\bar{n}\sub{d}}}
\safemath{\npf}{\ensuremath{\bar{n}\sub{p}}}
\safemath{\code}{\ensuremath{\mathcal{C}}}
\safemath{\err}{\ensuremath{\epsilon}}
\safemath{\rp}{\ensuremath{\rho\sub{p}}}
\safemath{\rd}{\ensuremath{\rho\sub{d}}}
\safemath{\cohtime}{\ensuremath{T\sub{c}}}
\safemath{\cohbw}{\ensuremath{B\sub{c}}}
\safemath{\nmax}{\ensuremath{\ell\sub{m}}}
\safemath{\ntot}{\ensuremath{n\sub{tot}}}
\safemath{\nul}{\ensuremath{n\sub{ul}}}
\safemath{\ndl}{\ensuremath{n\sub{dl}}}

\safemath{\yp}{\ensuremath{\randvecy_{\nu}^{(\text{p})}}}
\safemath{\yd}{\ensuremath{\randvecy_{\nu}^{(\text{d})}}}
\safemath{\ypd}{\ensuremath{\vecy_{\nu}^{(\text{p})}}}
\safemath{\ydd}{\ensuremath{\vecy_{\nu}^{(\text{d})}}}

\safemath{\ypf}{\ensuremath{\bar{\randvecy}_{\nu}^{(\text{p})}}}
\safemath{\ydf}{\ensuremath{\bar{\randvecy}_{\nu}^{(\text{d})}}}
\safemath{\ypdf}{\ensuremath{\bar{\vecy}_{\nu}^{(\text{p})}}}
\safemath{\yddf}{\ensuremath{\bar{\vecy}_{\nu}^{(\text{d})}}}

\safemath{\xp}{\ensuremath{\vecx^{(\text{p})}}}
\safemath{\xd}{\ensuremath{\randvecx_{\nu}^{(\text{d})}}}
\safemath{\xdd}{\ensuremath{\vecx_{\nu}^{(\text{d})}}}

\safemath{\xpf}{\ensuremath{\bar{\vecx}^{(\text{p})}}}
\safemath{\xdf}{\ensuremath{\bar{\randvecx}_{\nu}^{(\text{d})}}}
\safemath{\xddf}{\ensuremath{\bar{\vecx}_{\nu}^{(\text{d})}}}

\safemath{\xdb}{\ensuremath{\overline{\randvecx}^{(\text{d})}}}
\safemath{\Pxd}{\ensuremath{P_{\randvecx^{(\text{d})}}}}

\safemath{\xpbar}{\ensuremath{\overline{\matX}^{(\text{p})}}}
\safemath{\xdbar}{\ensuremath{\overline{\randmatX}^{(\text{d})}}}

\safemath{\xdv}{\ensuremath{\randvecx^{(\text{d})}}}
\safemath{\xdbarv}{\ensuremath{\overline{\randvecx}^{(\text{d})}}}
\safemath{\ydv}{\ensuremath{\randvecy^{(\text{d})}}}

\safemath{\xdr}{\ensuremath{\matX^{(\text{d})}}}

\safemath{\ttx}{\ensuremath{\tau\sub{tx}}}
\safemath{\trx}{\ensuremath{\tau\sub{rx}}}
\safemath{\ack}{\ensuremath{\mathrm{s}}}
\safemath{\nack}{\ensuremath{\mathrm{c}}}

\safemath{\mI}{\ensuremath{i\lro{\randvecy ; \randvecx}}} 				% i(Y;X)
\safemath{\randveca}{\bm{A}}
\safemath{\randvecb}{\bm{B}}
\safemath{\randvecc}{\bm{C}}
\safemath{\randvecd}{\bm{D}}
\safemath{\randvece}{\bm{E}}
\safemath{\randvecf}{\bm{F}}
\safemath{\randvecg}{\bm{G}}
\safemath{\randvech}{\bm{H}}
\safemath{\randveci}{\bm{I}}
\safemath{\randvecj}{\bm{J}}
\safemath{\randveck}{\bm{K}}
\safemath{\randvecl}{\bm{L}}
\safemath{\randvecm}{\bm{M}}
\safemath{\randvecn}{\bm{N}}
\safemath{\randveco}{\bm{O}}
\safemath{\randvecp}{\bm{P}}
\safemath{\randvecq}{\bm{Q}}
\safemath{\randvecr}{\bm{R}}
\safemath{\randvecs}{\bm{S}}
\safemath{\randvect}{\bm{T}}
\safemath{\randvecu}{\bm{U}}
\safemath{\randvecv}{\bm{V}}
\safemath{\randvecw}{\bm{W}}
\safemath{\randvecx}{\bm{X}}
\safemath{\randvecy}{\bm{Y}}
\safemath{\randvecz}{\bm{Z}}
\safemath{\randvecphi}{\bm{\Phi}}

\safemath{\randmatA}{\amsmathbb{A}}
\safemath{\randmatB}{\amsmathbb{B}}
\safemath{\randmatC}{\amsmathbb{C}}
\safemath{\randmatD}{\amsmathbb{D}}
\safemath{\randmatE}{\amsmathbb{E}}
\safemath{\randmatF}{\amsmathbb{F}}
\safemath{\randmatG}{\amsmathbb{G}}
\safemath{\randmatH}{\amsmathbb{H}}
\safemath{\randmatI}{\amsmathbb{I}}
\safemath{\randmatJ}{\amsmathbb{J}}
\safemath{\randmatK}{\amsmathbb{K}}
\safemath{\randmatL}{\amsmathbb{L}}
\safemath{\randmatM}{\amsmathbb{M}}
\safemath{\randmatN}{\amsmathbb{N}}
\safemath{\randmatO}{\amsmathbb{O}}
\safemath{\randmatP}{\amsmathbb{P}}
\safemath{\randmatQ}{\amsmathbb{Q}}
\safemath{\randmatR}{\amsmathbb{R}}
\safemath{\randmatS}{\amsmathbb{S}}
\safemath{\randmatT}{\amsmathbb{T}}
\safemath{\randmatU}{\amsmathbb{U}}
\safemath{\randmatV}{\amsmathbb{V}}
\safemath{\randmatW}{\amsmathbb{W}}
\safemath{\randmatX}{\amsmathbb{X}}
\safemath{\randmatY}{\amsmathbb{Y}}
\safemath{\randmatZ}{\amsmathbb{Z}}
\safemath{\randmatSigma}{\mathbb{\Sigma}}
\safemath{\randmatPhi}{\mathbb{\Phi}}
\safemath{\randmatLambda}{\mathbb{\Lambda}}

\safemath{\matSigma}{\bm{\Sigma}}
\safemath{\matPhi}{\bm{\Phi}}
\safemath{\matLambda}{\bm{\Lambda}}

\usepackage{glossaries}
\glsdisablehyper
\newglossaryentry{simo}
{
  name={SIMO},
  description={single-input multiple-output},
  first={\glsentrydesc{simo} (\glsentrytext{simo})}
}

\newglossaryentry{mmse}
{
  name={MMSE},
  description={minimum mean-square error},
  first={\glsentrydesc{mmse} (\glsentrytext{mmse})}
}

\newglossaryentry{lmmse}
{
  name={LMMSE},
  description={linear minimum mean-square error},
  first={\glsentrydesc{lmmse} (\glsentrytext{lmmse})}
}

\newglossaryentry{mf}
{
  name={MF},
  description={matched filtering},
  first={\glsentrydesc{mf} (\glsentrytext{mf})}
}

\newglossaryentry{qlmmse}
{
  name={QLMMSE},
  description={quasilinear MMSE},
  first={\glsentrydesc{qlmmse} (\glsentrytext{qlmmse})}
}

\newglossaryentry{na}
{
  name={NA},
  description={normal approximation},
  first={\glsentrydesc{na} (\glsentrytext{na})},
  plural={NAs},
  descriptionplural={normal approximations},
  firstplural={\glsentrydescplural{na} (\glsentryplural{na})}
}

\newglossaryentry{zf}
{
  name={ZF},
  description={zero-forcing},
  first={\glsentrydesc{zf} (\glsentrytext{zf})}
}

\newglossaryentry{rzf}
{
  name={RZF},
  description={regularized zero-forcing},
  first={\glsentrydesc{rzf} (\glsentrytext{rzf})}
}

\newglossaryentry{sir}
{
  name={SIR},
  description={signal-to-interference ratio},
  first={\glsentrydesc{sir} (\glsentrytext{sir})}
}

\newglossaryentry{snr}
{
  name={SNR},
  description={signal-to-noise ratio},
  first={\glsentrydesc{snr} (\glsentrytext{snr})}
}

\newglossaryentry{mse}
{
  name={MSE},
  description={mean-squared error},
  first={\glsentrydesc{mse} (\glsentrytext{mse})}
}
\newglossaryentry{mmmse}
{
  name={M-MMSE},
  description={multicell MMSE},
  first={\glsentrydesc{mmmse} (\glsentrytext{mmmse})}
}

\newglossaryentry{ls}
{
  name={LS},
  description={least-squares},
  first={\glsentrydesc{ls} (\glsentrytext{ls})}
}

\newglossaryentry{mr}
{
  name={MR},
  description={maximum ratio},
  first={\glsentrydesc{mr} (\glsentrytext{mr})}
}

\newglossaryentry{ue}
{
  name={UE},
  description={user equipment},
  first={\glsentrydesc{ue} (\glsentrytext{ue})},
  plural={UEs},
  descriptionplural={user equipments},
  firstplural={\glsentrydescplural{ue} (\glsentryplural{ue})}
}

\newglossaryentry{ap}
{
  name={AP},
  description={access point},
  first={\glsentrydesc{ap} (\glsentrytext{ap})},
  plural={APs},
  descriptionplural={access points},
  firstplural={\glsentrydescplural{ap} (\glsentryplural{ap})}
}

\newglossaryentry{cpu}
{
  name={CPU},
  description={central processing unit},
  first={\glsentrydesc{cpu} (\glsentrytext{cpu})},
  plural={CPUs},
  descriptionplural={central processing units},
  firstplural={\glsentrydescplural{cpu} (\glsentryplural{cpu})}
}

\newglossaryentry{bs}
{
  name={BS},
  description={base station},
  first={\glsentrydesc{bs} (\glsentrytext{bs})},
  plural={BSs},
  descriptionplural={base stations},
  firstplural={\glsentrydescplural{bs} (\glsentryplural{bs})}
}

\newglossaryentry{ostbc}
{
  name={OSTBC},
  description={orthogonal space-time block code},
  first={\glsentrydesc{ostbc} (\glsentrytext{ostbc})},
  plural={OSTBCs},
  descriptionplural={orthogonal space-time block codes},
  firstplural={\glsentrydescplural{ostbc} (\glsentryplural{ostbc})}
}

\newglossaryentry{biawgn}
{
  name={bi-AWGN},
  description={binary-input additive white Gaussian noise},
  first={\glsentrydesc{biawgn} (\glsentrytext{biawgn})}
}

\newglossaryentry{flnf}
{
  name={FLNF},
  description={fixed-length no-feedback},
  first={\glsentrydesc{flnf} (\glsentrytext{flnf})}
}

\newglossaryentry{crc}
{
  name={CRC},
  description={cyclic redundancy check},
  first={\glsentrydesc{crc} (\glsentrytext{crc})}
}

\newglossaryentry{bsc}
{
  name={BSC},
  description={binary symmetric channel},
  first={\glsentrydesc{bsc} (\glsentrytext{bsc})}
  plural={BSCs},
  descriptionplural={binary symmetric channels},
  firstplural={\glsentrydescplural{bsc} (\glsentryplural{bsc})}
}

\newglossaryentry{bec}
{
  name={BEC},
  description={binary-erasure channel},
  first={\glsentrydesc{bec} (\glsentrytext{bec})}
}
\newglossaryentry{awgn}
{
  name={AWGN},
  description={additive white Gaussian noise},
  first={\glsentrydesc{awgn} (\glsentrytext{awgn})}
}
\newglossaryentry{3gpp}
{
  name={3GPP},
  description={3rd generation partnership project},
  first={\glsentrydesc{3gpp} (\glsentrytext{3gpp})}
}

\newglossaryentry{dl}
{
  name={DL},
  description={downlink},
  first={\glsentrydesc{dl} (\glsentrytext{dl})}
}

\newglossaryentry{LED}
{
  name={LED},
  description={light emitting diode},
  first={\glsentrydesc{LED} (\glsentrytext{LED})},
  plural={LEDs},
  descriptionplural={light emitting diodes},
  firstplural={\glsentrydescplural{LED} (\glsentryplural{LED})}
}

\newglossaryentry{lte}
{
  name={LTE},
  description={long term evolution},
  first={\glsentrydesc{lte} (\glsentrytext{lte})}
}
\newglossaryentry{ofdm}
{
  name={OFDM},
  description={orthogonal frequency-division multiplexing},
  first={\glsentrydesc{ofdm} (\glsentrytext{ofdm})}
}

\newglossaryentry{ul}
{
  name={UL},
  description={uplink},
  first={\glsentrydesc{ul} (\glsentrytext{ul})}
}
\newglossaryentry{rb}
{
  name={RB},
  description={resource block},
  first={\glsentrydesc{rb} (\glsentrytext{rb})},
  plural={RBs},
  descriptionplural={resource blocks},
  firstplural={\glsentrydescplural{rb} (\glsentryplural{rb})}
}
\newglossaryentry{cu}
{
  name={CU},
  description={channel use},
  first={\glsentrydesc{cu} (\glsentrytext{cu})},
  plural={CUs},
  descriptionplural={channel uses},
  firstplural={\glsentrydescplural{cu} (\glsentryplural{cu})}
}
\newglossaryentry{tti}
{
  name={TTI},
  description={transmission time interval},
  first={\glsentrydesc{tti} (\glsentrytext{tti})},
  plural={TTI's},
  descriptionplural={transmission time intervals},
  firstplural={\glsentrydescplural{tti} (\glsentryplural{tti})}
}
\newglossaryentry{iid}
{
  name={i.i.d.},
  description={independent and identically distributed},
  first={\glsentrydesc{iid} (\glsentrytext{iid})}
}
\newglossaryentry{psd}
{
  name={PSD},
  description={power spectral density},
  first={\glsentrydesc{psd} (\glsentrytext{psd})}
}

\newglossaryentry{mimo}
{
  name={MIMO},
  description={multiple-input multiple-output},
  first={\glsentrydesc{mimo} (\glsentrytext{mimo})}
}

\newglossaryentry{bler}
{
  name={BLER},
  description={block error probability},
  first={\glsentrydesc{bler} (\glsentrytext{bler})}
}
\newglossaryentry{mtc}
{
  name={MTC},
  description={machine-type communication},
  first={\glsentrydesc{mtc} (\glsentrytext{mtc})}
}
\newacronym{csi}{CSI}{channel state information}
\newglossaryentry{dvbt}
{
  name={DVB-T},
  description={terrestrial digital video broadcasting},
  first={\glsentrydesc{dvbt} (\glsentrytext{dvbt})}
}
\newglossaryentry{pat}
{
  name={PAT},
  description={pilot-assisted transmission},
  first={\glsentrydesc{pat} (\glsentrytext{pat})}
}
\newglossaryentry{ml}
{
  name={ML},
  description={maximum likelihood},
  first={\glsentrydesc{ml} (\glsentrytext{ml})}
}
\newglossaryentry{dt}
{
  name={DT},
  description={dependence-testing},
  first={\glsentrydesc{dt} (\glsentrytext{dt})}
}
\newglossaryentry{ustm}
{
  name={USTM},
  description={unitary space-time modulation},
  first={\glsentrydesc{ustm} (\glsentrytext{ustm})}
}
\newglossaryentry{siso}
{
  name={SISO},
  description={single-input single-output},
  first={\glsentrydesc{siso} (\glsentrytext{siso})}
}
\newglossaryentry{snn}
{
  name={SNN},
  description={scaled nearest-neighbor},
  first={\glsentrydesc{snn} (\glsentrytext{snn})}
}

\newglossaryentry{snnd}
{
  name={SNND},
  description={scaled nearest-neighbor decoding},
  first={\glsentrydesc{snnd} (\glsentrytext{snnd})}
}
\newglossaryentry{rcus}
{
  name={RCUs},
  description={random-coding union bound with parameter $s$},
  first={\glsentrydesc{rcus} (\glsentrytext{rcus})}
}
\newglossaryentry{osd}
{
  name={OSD},
  description={ordered statistics decoding},
  first={\glsentrydesc{osd} (\glsentrytext{osd})}
}
\newglossaryentry{llr}
{
  name={LLR},
  description={log-likelihood ratio},
  first={\glsentrydesc{llr} (\glsentrytext{llr})}
   plural={LLR's},
  descriptionplural={log-likelihood ratios},
  firstplural={\glsentrydescplural{llr} (\glsentryplural{llr})}
}
\newglossaryentry{qpsk}
{
  name={QPSK},
  description={quaternary phase shift keying},
  first={\glsentrydesc{qpsk} (\glsentrytext{qpsk})}
}
\newglossaryentry{nlos}
{
  name={NLOS},
  description={non-line-of-sight},
  first={\glsentrydesc{nlos} (\glsentrytext{nlos})}
}
\newglossaryentry{los}
{
  name={LOS},
  description={line-of-sight},
  first={\glsentrydesc{los} (\glsentrytext{los})}
}
\newglossaryentry{mgf}
{
  name={MGF},
  description={moment-generating function},
  first={\glsentrydesc{mgf} (\glsentrytext{mgf})}
}
\newglossaryentry{cgf}
{
  name={CGF},
  description={cumulant-generating function},
  first={\glsentrydesc{cgf} (\glsentrytext{cgf})},
  plural={CGFs},
  descriptionplural={cumulant-generating functions},
  firstplural={\glsentrydescplural{cgf} (\glsentryplural{cgf})}
}
\newglossaryentry{pdf}
{
  name={PDF},
  description={probability density function},
  first={\glsentrydesc{pdf} (\glsentrytext{pdf})},
   plural={PDF's},
  descriptionplural={probability density functions},
  firstplural={\glsentrydescplural{pdf} (\glsentryplural{pdf})}
}

\newglossaryentry{ccdf}
{
  name={CCDF},
  description={complementary cumulative distribution function},
  first={\glsentrydesc{ccdf} (\glsentrytext{ccdf})}
}

\newglossaryentry{cdf}
{
  name={CDF},
  description={cumulative distribution function},
  first={\glsentrydesc{cdf} (\glsentrytext{cdf})}
}

\newglossaryentry{gmi}
{
  name={GMI},
  description={generalized mutual information},
  first={\glsentrydesc{gmi} (\glsentrytext{gmi})}
}

\newglossaryentry{arq}
{
  name={ARQ},
  description={automatic repeat request},
  first={\glsentrydesc{arq} (\glsentrytext{arq})}
}

\newglossaryentry{harq}
{
  name={HARQ},
  description={hybrid automatic repeat request},
  first={\glsentrydesc{harq} (\glsentrytext{harq})}
}

\newglossaryentry{fbl}
{
  name={FBL-NF},
  description={fixed blocklength with no feedback},
  first={\glsentrydesc{fbl} (\glsentrytext{fbl})}
}
\newglossaryentry{ssnf}
{
  name={SS-NF},
  description={single-shot no-feedback},
  first={\glsentrydesc{ssnf} (\glsentrytext{ssnf})}
}

\newacronym{urllc}{URLLC}{ultra-reliable low-latency communications}

\newglossaryentry{vlsf}
{
  name={VLSF},
  description={Variable-length stop-feedback},
  first={\glsentrydesc{vlsf} (\glsentrytext{vlsf})}
}

\newglossaryentry{vlnsf}
{
  name={VLNSF},
  description={variable-length noisy stop feedback},
  first={\glsentrydesc{vlnsf} (\glsentrytext{vlnsf})}
}

\newglossaryentry{dmt}
{
  name={DMT},
  description={diversity-multiplexing tradeoff},
  first={\glsentrydesc{dmt} (\glsentrytext{dmt})}
}

\newglossaryentry{dmdt}
{
  name={DMDT},
  description={diversity-multiplexing-delay tradeoff},
  first={\glsentrydesc{dmdt} (\glsentrytext{dmdt})}
}

\newglossaryentry{tddt}
{
  name={TDDT},
  description={throughput-diversity-delay tradeoff},
  first={\glsentrydesc{tddt} (\glsentrytext{tddt})}
}

\newglossaryentry{tdd}
{
  name={TDD},
  description={time-division duplex},
  first={\glsentrydesc{tdd} (\glsentrytext{tdd})}
}

\newglossaryentry{stbc}
{
  name={STBC},
  description={space-time block-codes},
  first={\glsentrydesc{stbc} (\glsentrytext{stbc})},
  plural={STBC's},
  descriptionplural={space-time block-codes},
  firstplural={\glsentrydescplural{stbc} (\glsentryplural{stbc})}
}

\newglossaryentry{harqir}
{
  name={HARQ-IR},
  description={hybrid automatic repeat request with incremental redundancy},
  first={\glsentrydesc{harqir} (\glsentrytext{harqir})}
}

\newglossaryentry{harqcc}
{
  name={HARQ-CC},
  description={hybrid automatic repeat request with chase combining},
  first={\glsentrydesc{harqcc} (\glsentrytext{harqcc})}
}

\newglossaryentry{wp1}
{
  name={w.p.1.},
  description={with probability one},
  first={\glsentrydesc{wp1} (\glsentrytext{wp1})}
}

\newglossaryentry{vlff}
{
  name={VLFF},
  description={variable-length full feedback},
  first={\glsentrydesc{vlff} (\glsentrytext{vlff})}
}

\newglossaryentry{dmc}
{
  name={DMC},
  description={discrete memoryless channel},
  first={\glsentrydesc{dmc} (\glsentrytext{dmc})}
  plural={DMCs},
  descriptionplural={discrete memoryless channels},
  firstplural={\glsentrydescplural{dmc} (\glsentryplural{dmc})}
}

\newglossaryentry{dnn}
{
  name={DNN},
  description={deep neural network},
  first={\glsentrydesc{dnn} (\glsentrytext{dnn})}
  plural={DNNs},
  descriptionplural={deep neural networks},
  firstplural={\glsentrydescplural{dnn} (\glsentryplural{dnn})}
}

\newglossaryentry{nn}
{
  name={NN},
  description={neural network},
  first={\glsentrydesc{nn} (\glsentrytext{nn})}
  plural={NNs},
  descriptionplural={neural networks},
  firstplural={\glsentrydescplural{nn} (\glsentryplural{nn})}
}

\newglossaryentry{cnn}
{
  name={CNN},
  description={convolutional neural network},
  first={\glsentrydesc{cnn} (\glsentrytext{cnn})}
  plural={CNNs},
  descriptionplural={convolutional neural networks},
  firstplural={\glsentrydescplural{cnn} (\glsentryplural{cnn})}
}

\newglossaryentry{ber}
{
  name={BER},
  description={bit error rate},
  first={\glsentrydesc{ber} (\glsentrytext{ber})}
}

\newglossaryentry{scss}
{
  name={SCSS},
  description={\emph{single-channel} source separation},
  first={\glsentrydesc{scss} (\glsentrytext{scss})}
}

\newglossaryentry{fft}
{
  name={FFT},
  description={fast Fourier transform},
  first={\glsentrydesc{fft} (\glsentrytext{fft})}
}

\newglossaryentry{soi}
{
  name={SOI},
  description={signal of interest},
  first={\glsentrydesc{soi} (\glsentrytext{soi})}
}

\newglossaryentry{map}
{
  name={MAP},
  description={maximum \emph{a posteriori}},
  first={\glsentrydesc{map} (\glsentrytext{map})}
}

\newglossaryentry{usc}
{
  name={TDC},
  description={``temporal-diversity" condition},
  first={\glsentrydesc{usc} (\glsentrytext{usc})}
}

\newglossaryentry{qam}
{
  name={QAM},
  description={quadrature amplitude modulation},
  first={\glsentrydesc{qam} (\glsentrytext{qam})}
}

\newglossaryentry{mit}
{
  name={MIT},
  description={Massachusetts Institute of Technology},
  first={\glsentrydesc{mit} (\glsentrytext{mit})}
}

\newglossaryentry{rf}
{
  name={RF},
  description={radio-frequency},
  first={\glsentrydesc{rf} (\glsentrytext{rf})}
}
\usepackage{pgfplots}
\pgfplotsset{compat=1.14}
\usepgfplotslibrary{groupplots}
\usetikzlibrary{pgfplots.groupplots} 
\usepackage{subcaption}
\usepackage{tabularx}
\usepackage{booktabs}

\usepackage{soul}   % in your preamble

\newtheorem{theorem}{Theorem}
\newtheorem{lemma}{Lemma}

\makeatletter
\newcommand\footnoteref[1]{\protected@xdef\@thefnmark{\ref{#1}}\@footnotemark}
\makeatother
\begin{document}
\title{Learning to Separate RF Signals Under Uncertainty: Detect-Then-Separate vs.~Unified Joint Models}

\author{\IEEEauthorblockN{Ariel Rodrigez$^{\dagger}$, Alejandro Lancho$^{\star}$ and Amir Weiss$^{\dagger}$}\\
\IEEEauthorblockA{
$^{\dagger}$Bar-Ilan University, Ramat Gan, Israel\\
$^{\star}$Universidad Carlos III de Madrid, Spain \& Gregorio Marañón Health Research Institute, Spain
%\\Emails: arielrod100@gmail.com, amir.weiss@biu.ac.il & alancho@ing.uc3m.es
}
% \\
% \begin{tabular}{ccc}
% $^{\dagger}$Faculty of Engineering & $^{\star}$Signal Theory and Communications Department\\
% Bar-Ilan University & Universidad Carlos III de Madrid %\\
% % arielrod100@gmail.com, amir.weiss@biu.ac.il & alancho@ing.uc3m.es
% \end{tabular}
% }
}
% \IEEEauthorblockA{
% Affiliations \\
% amir.weiss@biu.ac.il \vspace{-0.8cm}
% \thanks{Research was sponsored by ...} }  
 \maketitle

 \begingroup
\renewcommand\thefootnote{}\footnotetext{A. Lancho received funding from the Comunidad de Madrid under Grant Agreements No.~2023-T1/COM-29065 (César Nombela program), TEC-2024/COM-89, SYG-2024/COM-870, and from the Ministerio de Ciencia, Innovación y Universidades, Spain under Grant No.~PID2023-148856OA-I00.}
\addtocounter{footnote}{-1}
\endgroup
 
\begin{abstract}
The increasingly crowded radio frequency (RF) spectrum forces communication signals to coexist, creating heterogeneous interferers whose structure often departs from Gaussian models. Recovering the interference-contaminated signal of interest in such settings is a central challenge, especially in single-channel RF processing. Existing data-driven methods often assume that the interference type is known, yielding ensembles of specialized models that scale poorly with the number of interferers. We show that detect-then-separate (DTS) strategies admit an analytical justification: within a Gaussian mixture framework, a plug-in maximum \textit{a posteriori} detector followed by type-conditioned optimal estimation achieves asymptotic minimum mean-square error optimality under a mild temporal-diversity condition. This makes DTS a principled benchmark, but its reliance on multiple type-specific models limits scalability. Motivated by this, we propose a unified joint model (UJM), in which a single deep neural architecture learns to \emph{jointly} detect and separate when applied directly to the received signal. Using tailored UNet architectures for baseband (complex-valued) RF signals, we compare DTS and UJM on synthetic and recorded interference types, showing that a capacity-matched UJM can match oracle-aided DTS performance across diverse signal-to-interference-and-noise ratios, interference types, and constellation orders, including mismatched training and testing type-uncertainty proportions. These findings highlight UJM as a scalable and practical alternative to DTS, while opening new directions for unified separation under broader regimes.
\end{abstract}

\vspace{-0.2cm}
\section{Introduction}\label{sec:intro}

The modern radio-frequency (RF) spectrum has become a crowded and contested resource, shared by a multitude of technologies ranging from Wi-Fi and Bluetooth to ZigBee and emerging 5G systems~\cite{hu2018full}. This proliferation of wireless devices enables unprecedented connectivity, but also exacerbates the risk of interference, particularly in bands where distinct technologies overlap in both time and frequency. In such environments, the task of recovering a signal of interest (SOI) from an observed mixture of heterogeneous interfering signals is both practically urgent and theoretically challenging.

Traditional interference mitigation techniques often rely on spatial diversity (e.g., multiple antennas) or on carefully engineered signal structures that facilitate separation~\cite{kurt2023spatial}. Still, many practical scenarios are inherently single-channel, where neither array processing nor orthogonalization across time-frequency can be exploited. In these cases, the problem reduces to that of single-channel source separation (SCSS, e.g., \cite{tu2007particle,tu2008single}), a long-standing challenge that has recently gained renewed attention with the advent of data-driven methods.

Deep neural networks (DNNs) have shown promise in SCSS tasks across speech, music~\cite{luo2018tasnet,duan2008unsupervised}, and more recently RF signals~\cite{lancho2025rf}, provided that sufficiently rich training data is available. However, most existing approaches assume that the interference type is known in advance, and consequently train dedicated separation models tailored for each specific interferer. While this strategy can be effective in controlled conditions, it poses severe limitations, as it (i) requires maintaining a large ensemble of models; (ii) scales memory and training costs linearly with the number of interference types; and (iii) offers no guarantee of robustness under multiple interferers or interference-type uncertainty at run time.

This motivates the central question of the present work: \emph{Can one design a single data-driven model that successfully separates RF signals under interference-type uncertainty?} Addressing this question requires rethinking both theoretical underpinnings and practical modeling choices. On the one hand, it calls for a (generalized) statistical formulation that captures interference uncertainty and assesses the potential (sub)optimality of classical detection-separation cascades. On the other hand, it demands architectures that are not merely class-specific, but capable of \emph{jointly} detecting and mitigating multiple interference types from a single-channel mixture.

In this paper, we take a first step toward answering this question, and conjecture that the answer is affirmative. We formalize SCSS with interference-type uncertainty and study two complementary strategies. The first is a \emph{detect-then-separate} (DTS) approach, in which a classifier first identifies the interference type and then applies a specialized separator. The second is a \emph{unified joint model} (UJM), in which a single DNN is trained directly on mixtures from all classes and learns to implicitly detect, and to separate. Within a Gaussian mixture framework, we show that DTS admits an analytical justification: under a mild identifiability condition, a plug-in maximum \textit{a posteriori} (MAP) detector followed by class-specific minimum mean-square error (MMSE) estimator, which is linear in this case, achieves asymptotic optimality. This theoretical result establishes a principled baseline against which the performance of unified models can be assessed.

Building on this foundation, using the UNet architecture~\cite{ronneberger2015unet} adapted to complex-valued RF signals, we revisit design modifications that accommodate the statistical structure of baseband mixtures. We then compare the two strategies empirically on synthetic and recorded interferences from standardized datasets~\cite{rfchallenge}, but unlike prior works, considering SOIs of increasing constellation order. Our results demonstrate that while a lightweight UJM may fall short in challenging regimes, a capacity-matched UJM matches an oracle-aided DTS ensemble in terms of both waveform fidelity (i.e., MSE) and bit error rate (BER), all while requiring only a single set of parameters.

Our main contributions in this paper are as follows:
\begin{itemize}[itemsep=0.4pt,topsep=2pt,parsep=0.3pt,partopsep=0.4pt]
\item We formalize SCSS with interference-type uncertainty and derive an MMSE decomposition that clarifies the role of interference detection in separation. This formulation enables us to prove that a plug-in DTS strategy is asymptotically MMSE-optimal in an analytically tractable Gaussian mixture model.
\item We propose a unified approach, dubbed UJM, enabling joint detection and separation in a single learned model.
\item We validate our framework through extensive simulation experiments with multiple interference types and higher-order modulations, demonstrating that unified models can achieve the same accuracy as DTS, thus paving the way towards foundation-like models for source separation.%demonstrating when unified models suffice and when explicit detection remains advantageous.
\end{itemize}

\textit{Notation:} We use lowercase letters with standard font and sans-serif font, e.g., $x$ and $\rndx$, to denote deterministic and random scalars, respectively. Similarly, we use $\vecx$ and $\rvecx$ for deterministic and random vectors, respectively; and $\matX$ and $\rmatX$ for deterministic and random matrices, respectively. The $n$-th random sample of the random vector $\rvecx$ is denoted by $\rndx[n]$. For $K\in\naturals$, we denote the set $\setT_K \triangleq \{1, \ldots, K\}$. The probability of an event $\setA$ is denoted by $\Prob(\setA)$. For brevity, we refer to the complex normal distribution as Gaussian. We denote $\matC_{\rndz\rndw} \triangleq \Ex{}{ \rvecz\herm{\rvecw}} \in \complexset^{N_{\rndz} \times N_{\rndw}}$ as the covariance matrix of $\rvecz \in \complexset^{N_\rndz \times 1}$ and $\rvecw \in \complexset^{N_\rndw \times 1}$ (specializing to $\mathbf{C}_{\rndz}$ for $\rvecz = \rvecw$).\vspace{-0.3cm}

\section{Problem Formulation}\label{sec:problem}\vspace{-0.2cm}
We consider a single-channel receiver that observes a superposition of two signals: a desired SOI contaminated by an interfering signal. Unlike in classical interference rejection, where the interference structure is fixed and known, here it is \emph{random}. In other words, while the receiver always observes only a single mixture, there is uncertainty regarding which interference type has been realized. Formally, consider the following baseband signal model
{\setlength{\belowdisplayskip}{4pt} \setlength{\belowdisplayshortskip}{4pt}
    \setlength{\abovedisplayskip}{4pt} \setlength{\abovedisplayshortskip}{4pt}
% \begin{equation}
%     \rndy[n] = \rnds[n] + \frac{1}{\sqrt{\rho_{\text{\tiny SIR}}}}\cdot\rndv_{\rndk}[n]+\frac{1}{\sqrt{\rho_{\text{\tiny SNR}}}}\cdot\rndw[n],
%     \label{eq:1}
% \end{equation}}
\begin{equation}
    \rndy[n] = \rnds[n] + \rho_{\text{\tiny SIR}}^{-\frac{1}{2}}\cdot\rndv_{\rndk}[n]+\rho_{\text{\tiny SNR}}^{-\frac{1}{2}}\cdot\rndw[n],
    \label{eq:1}
\end{equation}}
\par\noindent where
\begin{itemize}
    \item $\rnds[n] \in \complexset$ denotes the SOI, generated according to a known protocol (e.g., OFDM~\cite{hwang2008ofdm});
    \item $\rndv_{\rndk}[n] \in \complexset$ represents the interfering signal, drawn from a finite set of $K\in\naturals$ possible interference types, indexed by the random variable $\rndk\in\setT_k$, denoting its type, i.e.,
    {\setlength{\belowdisplayskip}{4pt} \setlength{\belowdisplayshortskip}{4pt}
    \setlength{\abovedisplayskip}{4pt} \setlength{\abovedisplayshortskip}{4pt}
    \begin{equation}
        \Prob\left(\rndv_{\rndk}[n]=\rndv_{k}[n]\right) = p_k\in[0,1] \; \Rightarrow \; \sum_{k\in\setT_K}p_k=1;
    \end{equation}}
% \par\noindent such that $\sum_{k\in\setT_K}p_k=1$;
    \item $\rndw[n]\in\complexset$ is the channel additive white Gaussian noise, statistically independent of both $\rnds[n]$ and $\rndv_{\rndk}[n]$; and
    \item $\rho_{\text{\tiny SIR}}, \rho_{\text{\tiny SNR}} \in \reals_+$ denote the signal-to-interference ratio (SIR) and signal-to-noise ratio (SNR), respectively.
\end{itemize}
For simplicity, and to facilitate the definitions of $\rho_{\text{\tiny SIR}}$ and $ \rho_{\text{\tiny SNR}}$, we further assume that the $\rnds[n], \rndv_{\rndk}[n]$ and $\rndw[n]$ are all unit variance. Furthermore, in this work we focus exclusively on the low SIR regime, and further assume that $\rho_{\text{\tiny SIR}}\ll \rho_{\text{\tiny SNR}}$. Thus, we henceforth refer to the ``effective interference", define as $\rndb_\rndk[n]\triangleq\rho^{-\frac{1}{2}}_{\text{\tiny SIR}}\rndv_{\rndk}[n]+\rho^{-\frac{1}{2}}_{\text{\tiny SNR}}\rndw[n]$, for which $\Varop\left(\rndb_\rndk[n]\right)\approx1/\rho_{\text{\tiny SIR}}$.

Now, consider a batch of $N\in\naturals$ samples, and let $\rvecs \triangleq \tp{[\rnds[1] \cdots \rnds[N]]}$ and $\rvecb_{\rndk} \triangleq \tp{[\rndb_{\rndk}[1] \cdots \rndb_{\rndk}[N]]}$. Using these notations, for $N$ samples, \eqref{eq:1} may be compactly written as
{\setlength{\belowdisplayskip}{4pt} \setlength{\belowdisplayshortskip}{4pt}
    \setlength{\abovedisplayskip}{4pt} \setlength{\abovedisplayshortskip}{4pt}
\begin{equation}
    \rvecy = \rvecs + \rvecb_{\rndk}\in\complexset^{N\times1}.
    \label{eq:2}
\end{equation}
}
% For a specific realization of the interference type, say $k \in \setT_K$, we denote by $\rvecb_k \triangleq \rvecb_{\rndk}\mid \rndk=k$ the corresponding interference vector, and by $\rvecy_k \triangleq \rvecy \mid \rndk=k$ the received mixture containing it. The task of the receiver is then to recover $\rvecs$ from the observation $\rvecy$. 
\par\noindent In the general formulation of SCSS, now extended to account for interference-type uncertainty, we seek an estimator $\widehat{\rvecs}$ that minimizes the expected loss with respect to a prescribed loss function $\ell:\complexset^{N\times1}\times\complexset^{N\times1}\to\positivereals$, namely
{\setlength{\belowdisplayskip}{4pt} \setlength{\belowdisplayshortskip}{4pt}
    \setlength{\abovedisplayskip}{4pt} \setlength{\abovedisplayshortskip}{4pt}
\begin{equation}\label{eq:expectedloss}
    g^* = \arg\min_{g:\complexset^{N\times1}\to\complexset^{N\times1}} \Exop\left[\ell(\rvecs,g(\rvecy))\mid \rvecy\right]\; \Rightarrow \; \widehat{\rvecs}= g^*(\rvecy).
\end{equation}}
\par\noindent In the data-driven formulation of this problem, we assume that a dataset of $D\in\naturals$ independent, identically distributed (i.i.d.) copies of baseband mixture-SOI pairs is available, denoted by $\{(\rvecy^{(i)}, \rvecs^{(i)})\}_{i=1}^D$, whose real and imaginary parts are their in-phase and quadrature components, respectively. In this case, we seek to \emph{learn}, based on the dataset, an estimator $\widehat{\rvecs}$ that minimizes the empirical loss, namely
{\setlength{\belowdisplayskip}{3pt} \setlength{\belowdisplayshortskip}{3pt}
    \setlength{\abovedisplayskip}{3pt} \setlength{\abovedisplayshortskip}{3pt}
\begin{align}
    \hspace{-0.05cm}\vectheta^* = \arg\min_{\vectheta\in\reals^{N_{\theta}\times1}} \sum_{i=1}^D\hspace{-0.025cm}\ell\left(\rvecs^{(i)},g_{\vectheta}\left(\rvecy^{(i)}\right)\right) 
    \Rightarrow \; \widehat{\rvecs}\triangleq g_{\vectheta^*}(\rvecy),\label{eq:separationforgeneralizedsscs}
    % \vectheta^* &= \arg\min_{\vectheta\in\reals^{N_{\theta}\times1}} \sum_{i=1}^D\ell\left(\rvecs^{(i)},g_{\vectheta}\left(\rvecy^{(i)}\right)\right) \label{eq:empiricaloss}\\
    % \Rightarrow \; \widehat{\rvecs}&\triangleq g_{\vectheta^*}(\rvecy),\label{eq:separationforgeneralizedsscs}
\end{align}}
\par \noindent where $\vectheta\in\reals^{N_{\theta}\times1}$ is a high-dimensional vector of parameters (i.e., degrees of freedom), such as the weights of a DNN, with $N_{\theta}\in\naturals$ denoting its dimension, and $g_{\vectheta}(\cdot)$ is the corresponding parametric function (e.g., a DNN). In this generalized formulation of SCSS, $\widehat{\rvecs}$ in \eqref{eq:separationforgeneralizedsscs} must identify (explicitly or implicitly) the type of interference and then apply a separator specifically tailored to the identified interference type.

\vspace{-0.2cm}
\section{SCSS Data-Driven Approaches with Interference Uncertainty}\label{sscs_approaches}
\vspace{-0.1cm}
\begin{figure*}
    \centering
    \begin{subfigure}{0.48\textwidth}
    \centering \includegraphics[width=\linewidth]{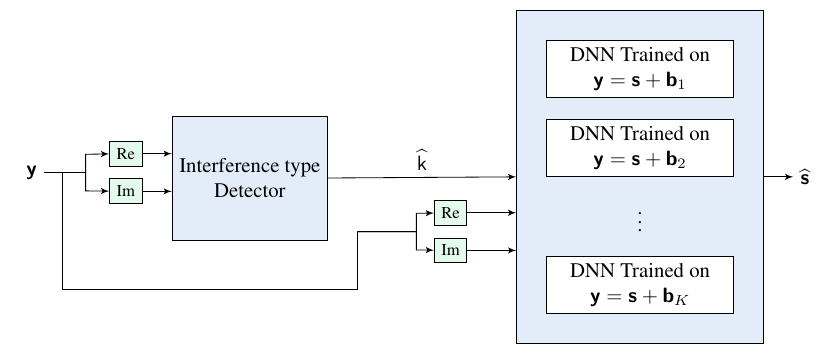}
     \caption{System model of the DTS approach, with an explicit detection block prior to the per-interference-type separation DNN blocks.}
     \label{fig:naive-approach}
    \end{subfigure}\hfill
    \begin{subfigure}{0.48\textwidth}
            \centering
            \includegraphics[width=0.75\linewidth]{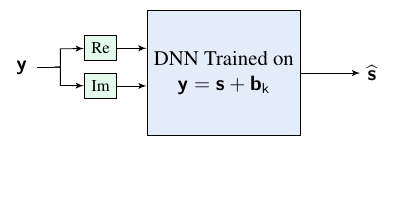}
            \caption{System model of the proposed UJM approach, with a single DNN block to perform joint detection and separation.}
            \label{fig:desiredDNN}
    \end{subfigure}
    \caption{Block diagrams of the two DNN architectures for signal separation under interference uncertainty.}\vspace{-0.35cm}
\end{figure*}

When the interference type is uncertain, standard data-driven source separation models, typically tailored to a single interference type, must be adapted. Two main strategies arise: the DTS approach, which trains a separate model per interference type and relies on a detector at runtime, and the UJM approach, which uses a \emph{single} DNN trained directly on unconditioned mixtures. Both approaches can be interpreted through the lens of the MMSE benchmark, but they differ in scalability and design philosophy. To make this connection precise, we first recall that the MMSE estimator is defined as
\begin{equation}
\widehat{\rvecs}_{\text{\tiny MMSE}} \triangleq \Exop\left[\rvecs \mid \rvecy\right].\label{eq:mmseestdef}
\end{equation}
 Indeed, by simple application of the law of total expectation, \eqref{eq:mmseestdef} can be expressed as
 {\setlength{\belowdisplayskip}{1pt} \setlength{\belowdisplayshortskip}{1pt}
    \setlength{\abovedisplayskip}{1pt} \setlength{\abovedisplayshortskip}{1pt}
\begin{align}
    % \widehat{\rvecs}_{\text{\tiny MMSE}} &= \Exop\left[\Exop\left[\rvecs \mid \rvecy, \rndk\right] \mid \rvecy\right] =\sum_{k=1}^{K}\Prob(\rndk=k\mid \rvecy)\Exop\left[\rvecs | \rvecy, \rndk=k\right] \\
    % &= \sum_{k=1}^{K}\Prob(\rndk=k\mid \rvecy) \widehat{\rvecs}_{\text{\tiny MMSE}}(k), \label{eq:5}
    \widehat{\rvecs}_{\text{\tiny MMSE}} &= \Exop\left[\Exop\left[\rvecs \mid \rvecy, \rndk\right] \mid \rvecy\right] = \sum_{k=1}^{K}\Prob(\rndk\!=\!k\mid \rvecy) \widehat{\rvecs}_{\text{\tiny MMSE}}(k), \label{eq:mmsedecompose}
\end{align}}
\par \noindent where $\widehat{\rvecs}_{\text{\tiny MMSE}}(k) \triangleq \Ex{}{\rvecs | \rvecy, \rndk=k}=\Ex{}{\rvecs | \rvecy_k}$ and we further denote $\rvecy_k \triangleq \rvecs + \rvecb_k$ given that $\rndk=k$. Thus, the MMSE estimator is a weighted average of all $K$ optimal estimators that are \emph{conditioned} on the interference type, with the weights being the posterior probabilities of $\rndk$ given $\rvecy$. This decomposition provides a useful perspective on the approaches discussed next.\vspace{-0.2cm}

\subsection{The Detect-Then-Separate Approach}

A natural way to handle interference uncertainty is to train $K$ separate DNN models, each specialized for one interference type. At runtime, a detector first classifies the interference type present in the received mixture, and the corresponding model is then applied to estimate the SOI. A block diagram of this detection-separation pipeline is shown in Fig.~\ref{fig:naive-approach}.

Formally, if $\{(\rvecy_k^{(i)}, \rvecs^{(i)})\}_{i=1}^D$ denotes a dataset of mixtures conditioned on interference type $k$, then the $k$-th DNN is trained to minimize
 {\setlength{\belowdisplayskip}{4pt} \setlength{\belowdisplayshortskip}{4pt}
    \setlength{\abovedisplayskip}{4pt} \setlength{\abovedisplayshortskip}{4pt}
\begin{equation}
    \vectheta_k^* = \arg\min_{\vectheta\in\reals^{N_{k}\times1}} \sum_{i=1}^D 
    \ell\left(\rvecs^{(i)}, g_{\vectheta}\left(\rvecy_k^{(i)}\right)\right),
\end{equation}}
\par\noindent yielding a collection of type-specific estimators $\{g_{\vectheta_k^*}\}_{k=1}^K$.

This approach mirrors the MMSE decomposition (\ref{eq:mmsedecompose}): the conditional estimators $g_{\vectheta_k^*}$ play the role of type-conditioned MMSE estimators, while the detector plays the role of selecting the most probable type, i.e., the one maximizing the posterior probability of $\rndk\!=\!k$ given the observation $\rvecy$. As we show in Section~\ref{subsec:analyticalgaussian}, under a mild temporal-diversity condition, the detector can be made asymptotically reliable, with which DTS can approach MMSE performance and serve as a nearly-optimal benchmark. Its main drawback, however, is scalability: the parameter count grows linearly with $K$, and the approach requires labeled data for each interference type.\vspace{-0.05cm}

\subsection{The Unified Joint Model Approach}\vspace{-0.05cm}
An alternative to DTS is to train a \emph{single} DNN directly on unconditioned mixtures, \emph{without} explicit knowledge of the interference type. In this unified joint model, the network implicitly learns to both infer the interference type and separate the SOI within a single architecture. This approach (Fig.~\ref{fig:desiredDNN}):
\begin{itemize}
    \item avoids the linear growth in parameter count with $K$; and
    \item eliminates the need for type-specific labeled datasets, offering a more scalable, practical solution.
\end{itemize}  

From the MMSE perspective, the UJM can be seen as attempting to approximate the MMSE estimator \eqref{eq:mmseestdef}, without explicitly decomposing it into type-conditioned components. The challenge, however, is that a lightweight unified model may struggle to match the performance of an ensemble of specialized models, particularly when $K$ is large or interference types are highly heterogeneous. In such regimes, increased model capacity or more sophisticated architectures may be required for the UJM to approach MMSE performance.

\subsection{Analytical Analysis for the Gaussian Mixture Signal Model}\label{subsec:analyticalgaussian}

To better understand the relations between DTS, UJM, and the MMSE benchmark, we now turn to a tractable Gaussian setting. This idealized model does not capture all practical signal structures, but it provides a clear analytical lens through which the role of interference uncertainty can be examined.

Specifically, suppose that $\rvecs \sim \jpg(\mathbf{0}, \bC_{\rnds})$ is a zero-mean Gaussian vector and, conditioned on $\rndk\!=\!k$, the interference satisfies $\rvecb_{k} \sim \jpg(\mathbf{0}, \bC_{\rndb_k})$. In this case, the received mixture
\begin{equation}
    \rvecy_k \sim \jpg\left(\mathbf{0}, \bC_{\rnds}+\bC_{\rndb_k}\right)
\end{equation}
is itself Gaussian, and the conditional MMSE estimator coincides with the linear MMSE (LMMSE) estimator, given by
\begin{equation}\label{eq:complexnormaly}
    \widehat{\rvecs}_{\text{\tiny LMMSE}}(k) 
    \triangleq \bC_{\rnds \rndy_k}\inv{\bC_{\rndy_k}}\rvecy_k 
    = \bC_{\rnds}\inv{\left(\bC_{\rnds}+\bC_{\rndb_k}\right)}\rvecy_k.
\end{equation}
By substituting these type-conditioned estimators into the MMSE decomposition \eqref{eq:mmsedecompose}, we obtain
 {\setlength{\belowdisplayskip}{4pt} \setlength{\belowdisplayshortskip}{4pt}
    \setlength{\abovedisplayskip}{4pt} \setlength{\abovedisplayshortskip}{4pt}
\begin{equation}
    \widehat{\rvecs}_{\text{\tiny MMSE}} 
    = \sum_{k=1}^{K} \Prob(\rndk=k \mid \rvecy)\,
      \widehat{\rvecs}_{\text{\tiny LMMSE}}(k),\label{eq:mmseislmmse}
\end{equation}}
\par\noindent i.e., the MMSE is a weighted average of $K$ linear estimators. In this case, if the interference type was known, only the corresponding conditional \emph{linear} estimator $\widehat{\rvecs}_{\text{\tiny LMMSE}}(k)$ would be required. More generally, if the MAP detector $\widehat{\rndk}_{\text{\tiny MAP}}$ of $\rndk$ reliably identifies the correct type, the quasi-linear\footnote{While the type-conditioned estimators $\{\widehat{\rvecs}_{\text{\tiny LMMSE}}(k)\}$ are linear with $\rvecy$, the weights $\{\Prob(\rndk=k \mid \rvecy)\}$ are certainly not, hence the \emph{quasi}-linearity.} MMSE plug-in (suboptimal) estimator, which is a DTS realization, given by
\begin{equation}\label{eq:mapqlmmse}
    \widehat{\rvecs}_{\text{\tiny DTS-GM}}
    \triangleq \widehat{\rvecs}_{\text{\tiny LMMSE}}\left(\widehat{\rndk}_{\text{\tiny MAP}}\right),
\end{equation}
can be shown---under a mild temporal-diversity condition---to asymptotically achieve the performance of the exact MMSE estimator. This formally justifies DTS as an asymptotically optimal strategy in this Gaussian mixture case, and potentially in more general cases. At the same time, it clarifies the challenge for the more attractive UJM architecture: to approximate the same (unconditioned) MMSE \emph{without} explicit type separation.

To formalize the claim of asymptotic optimality of DTS in this case, we first recall the ``temporal-diversity" condition (TDC)~\cite[Def.~1]{proofs}, under which optimal detection is increasingly accurate.\footnote{This is true even for a judiciously constructed sub-optimal detection.} For completeness, we revisit its definition.

\textit{Definition 1 (TDC):} Let $k\in\setT_K$ and let $\psi_N(\rvecy,k) \triangleq \frac{1}{N}\herm{\rvecy}\inv{\bC}_{\rndy_k} \rvecy-1$. The TDC is satisfied if, given $\rndk=k$, there does not exist $\ell \in \setT_{K} \backslash k$ for which $|\psi_N(\rvecy,\ell)|\xrightarrow[N\rightarrow\infty]{\text{a.s.}}0$.

Using the TDC, we recall the following key lemma, whose proof outline is given in Appendix~\ref{app:lemma1} for clarity.

\begin{lemma}\label{lem:TDC}~\cite[Lemma~1]{proofs}
Under the TDC, for any finite $\alpha>0$ independent of $N$,
{\setlength{\belowdisplayskip}{4pt} \setlength{\belowdisplayshortskip}{4pt}
 \setlength{\abovedisplayskip}{4pt} \setlength{\abovedisplayshortskip}{4pt}
\begin{equation}
    \Prob\left(\widehat{\rndk}_{\text{\tiny MAP}} \ne \rndk\right) 
    = o\left(\frac{1}{N^\alpha}\right).
\end{equation}}
\end{lemma}

% \noindent\textit{Lemma 1:} Under the TDC, for any finite $\alpha \in \positivereals$ independent of $N$,
% \begin{equation}
%     \Prob\left(\widehat{\rndk}_{\text{\tiny MAP}} \ne \rndk\right) = o\left(\frac{1}{N^\alpha}\right).
% \end{equation}

Consequently, when the interference type is asymptotically detectable, the DTS strategy achieves the MMSE performance in the large-$N$ limit, as established by the following theorem.
\begin{theorem}\textit{(DTS asymptotic optimality in the Gaussian mixture signal model)}\label{thm:DTS_MMSE}
Let
{\setlength{\belowdisplayskip}{4pt} \setlength{\belowdisplayshortskip}{4pt}
 \setlength{\abovedisplayskip}{4pt} \setlength{\abovedisplayshortskip}{4pt}
\begin{align}
    \varepsilon_{\text{\tiny MMSE}}^2(N) &\triangleq 
        \Exop\!\left[\left\| \widehat{\rvecs}_{\text{\tiny MMSE}} - \rvecs \right\|_2^2 \right], \\
    \varepsilon_{\text{\tiny DTS-GM}}^2(N) &\triangleq 
        \Exop\!\left[\left\| \widehat{\rvecs}_{\text{\tiny DTS-GM}} - \rvecs \right\|_2^2 \right],
\end{align}}
\par\noindent be the MSEs of the MMSE \eqref{eq:mmseislmmse} and DTS \eqref{eq:mapqlmmse} estimators, respectively, as functions of the sample size $N$. Then, under the TDC, for the Gaussian mixture signal model \eqref{eq:complexnormaly},
 {\setlength{\belowdisplayskip}{4pt} \setlength{\belowdisplayshortskip}{4pt}
    \setlength{\abovedisplayskip}{4pt} \setlength{\abovedisplayshortskip}{4pt}
\begin{equation}
    \lim_{N\to\infty} 
    \frac{\varepsilon_{\text{\tiny MMSE}}^2(N)}{\varepsilon_{\text{\tiny DTS-GM}}^2(N)} = 1.
\end{equation}}
\end{theorem}

Due to space considerations, the proof is omitted; see~\cite[Appendix B]{proofs} for a similar analysis. This result establishes that DTS can be asymptotically optimal (at least) in the Gaussian mixture setting, thereby providing a rigorous foundation for its use as a benchmark in the more general, data-driven setting.

\vspace{-0.1cm}
\section{Simulation Results}\label{sec:simulations}\vspace{-0.1cm}
In this section, we present simulation experiments for comparing the performance of DTS against the proposed UJM. The DTS approach assumes that the interference type is provided by an oracle, i.e., at the separation stage, the true interference type is assumed to be known. In practice, this detection task would have to be performed by a dedicated classifier (e.g., a pretrained DNN), and thus constitutes an additional source of potential errors. Here, the oracle assumption allows us to isolate the effect of the separation stage, and to assess the potential of unified models versus\ an even stronger competitor with an ideal detector.

For the separation task, both DTS and UJM rely on DNNs based on the UNet architecture, previously shown to be effective for RF signal separation problems (see, e.g., \cite{lee2022exploiting,proofs,lancho2025rf}). We emphasize that here, DTS is a DNN and not \eqref{eq:mapqlmmse}. Our goal in this section is not to optimize architectures for optimal performance, but rather to establish a proof of concept. Hence, we adopt a baseline UNet variant and restrict our attention to a small number of interference types. Specifically, we focus on the case $K=2$, recalling that our general formulation in Section~\ref{sscs_approaches} extends naturally to arbitrary~$K$.

The main architectural difference between DTS and UJM lies in the network depth. The DTS separator employs a UNet with five downsampling blocks (with the corresponding symmetric upsampling blocks), while the UJM requires an extended variant with eight downsampling/upsampling blocks. This additional depth is to (at least conceptually) compensate for the absence of an external detector and to match the performance of the DTS ensemble with a single network. To show this, we also compare a ``lightweight capacity" UJM with only five downsampling/upsampling blocks, denoted as 5L-UJM. Table~\ref{tab:unet_comparison} summarizes the main differences between the two approaches, as implemented here. Recall that, since DTS instantiates one network per possible interference class, its parameter count scales linearly with~$K$. Moreover, we note that we do not take into account in this comparison the additional parameters DTS would need for a detector.\vspace{-0.2cm}

\subsection{Simulation Setup}

We follow the probabilistic formulation in Section~\ref{sec:problem}, where each mixture instance contains either interference type~1 or type~2, selected according to a Bernoulli random variable with parameter~$p$, i.e., $\Prob\left(\rvecb_\rndk = \rvecb_{1}\right)=p$, and 
$\Prob\left(\rvecb_\rndk = \rvecb_{2}\right)=1-p$.

Within each mixture case, the SOI is fixed and generated as an $M$-ary phase-shift keying ($M$-PSK) waveform, shaped by a root-raised cosine filter with roll-off factor~0.5. Each frame spans 8~symbols, oversampled by a factor of~16, and includes an initial offset of 8~samples. For additional details, see \cite{papergithub}.

\begin{table}[t]
\centering
\caption{Comparison of DTS and UJM architectures. In ``\# Parameters", percentages are relative to the UNet from~\cite{rfchallenge}.}
\label{tab:unet_comparison}
\begin{tabularx}{\columnwidth}{lXX}
\toprule
 & \textbf{DTS} & \textbf{UJM} \\
\midrule
Detection & External oracle & Implicit \\
Architecture (UNet) & 5 down/5 up blocks & 8 down/8 up blocks \\
\# Parameters ($K=2$) & 8,445,396 (200\%) & 8,515,785 ($\approx$201.7\%) \\
\bottomrule
\end{tabularx}\vspace{-0.4cm}
\end{table}

The interference signals $\rvecb_\rndk$ are taken from the RF Challenge dataset \cite{rfchallenge}, which includes four types of interference signal recordings, for which the generation process is unknown. These signals are denoted as \emph{EMISignal1}, \emph{CommSignal2}, \emph{CommSignal3}, and \emph{CommSignal5G1}, but for brevity they are referred to here as \emph{EMIS1}, \emph{CS2}, \emph{CS3}, and \emph{CS5G1}, respectively. For a complete description of the interference dataset, see \cite{lancho2025rf}. For $K=2$, we therefore obtain six possible interference mixtures, which we denote as $\rvecb_1/\rvecb_2$ (e.g., CS2/CS3). Each interference frame is randomly cropped to length $N=40{,}960$, scaled to achieve a target effective interference level consistent with the desired SIR, and rotated by a random phase to emulate RF impairments. Since the recorded interference signals already contain additive noise, the scaling factor is computed in terms of signal-to-interference-plus-noise ratio (SINR), which in practice coincides with the effective SIR defined in Section~\ref{sec:problem}.

Datasets are formed by generating $D$ i.i.d.~pairs $\{(\rvecy^{(i)}, \rvecs^{(i)})\}_{i=1}^D$, with in-phase and quadrature components treated as real-valued channels. All networks are trained on datasets of equal nominal size, although in the DTS case this effectively doubles the total number of training samples due to its class-specific separation blocks. Performance is evaluated in terms of both waveform fidelity (MSE) and message recovery (BER), averaged over $2500$ realizations per interference level, with SINR swept from $-30$~dB to $0$~dB in steps of $3$~dB.

\vspace{-0.2cm}
\subsection{Numerical Results}
\begin{figure}[t]
    \centering
    \includegraphics[width=0.9\linewidth]{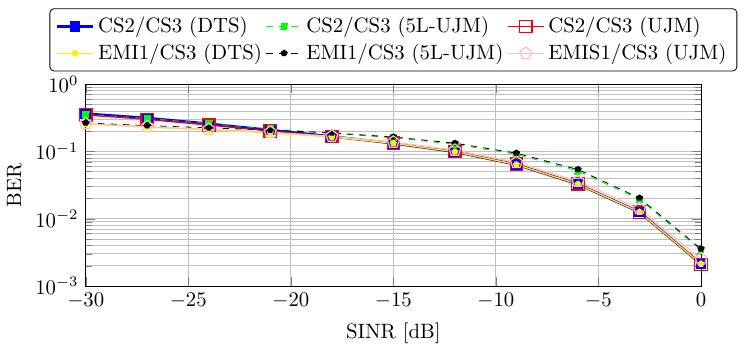}\vspace{-0.3cm}
    \caption{QPSK SOI with mixtures containing CS3 ($p=1/2$): BER vs. SINR for DTS, 5L-UJM and UJM.}
    \label{fig:qpsk_cs3_mixtures}\vspace{-0.2cm}
\end{figure}

\begin{figure}[t]
    \centering
    \includegraphics[width=0.9\linewidth]{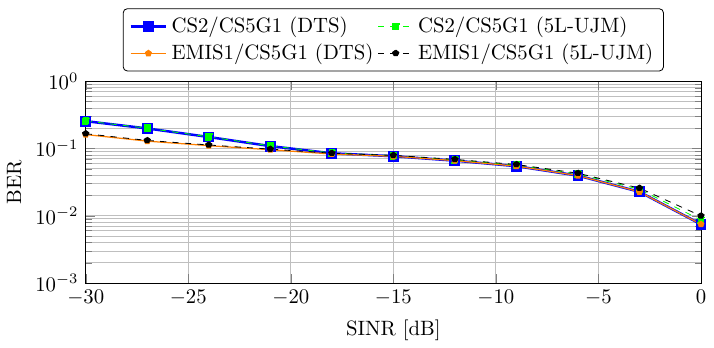}\vspace{-0.3cm}
    \caption{QPSK SOI with mixtures containing CS5G1 ($p=1/2$): BER vs. SINR for DTS, 5L-UJM and UJM.}
    \label{fig:qpsk_cs5g1_mixtures}\vspace{-0.5cm}
\end{figure}

%Figure~\ref{fig:qpsk+emi1ORcs2} illustrates the case EMISignal1/CommSignal2 interference, where UJM closely tracks DTS across the full SINR range, where the slight BER gaps are possibly caused by the averaging of their marginal-case performances.

\begin{figure*}
    \centering
    \includegraphics[width=0.975\linewidth]{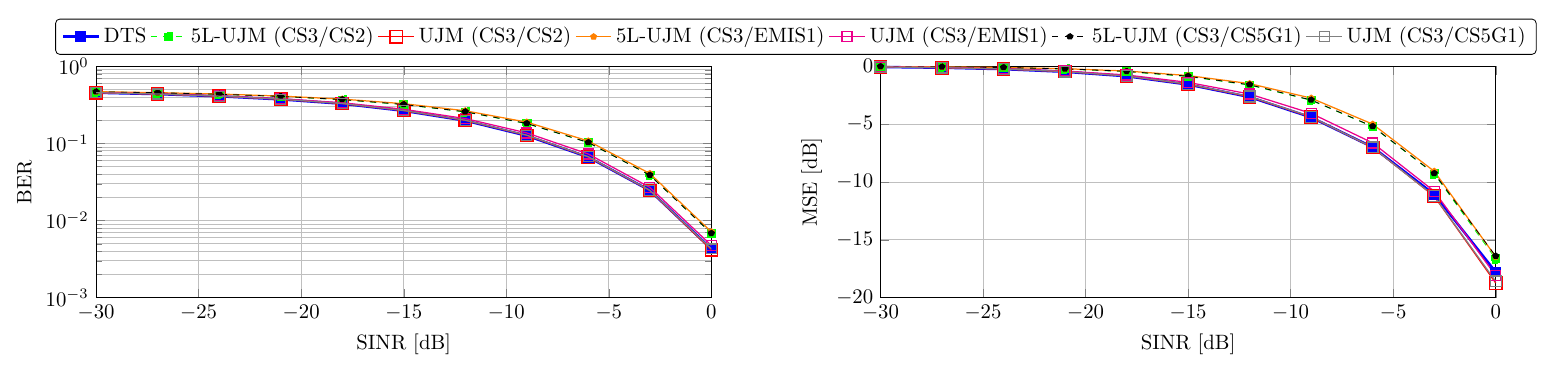}\vspace{-0.3cm}
    \caption{QPSK SOI with CS3 interference only ($p=1$): BER and MSE vs. SINR for DTS, 5L-UJM and UJM.}
    \label{fig:qpsk+cs3}\vspace{-0.4cm}
\end{figure*}
We now turn to the evaluation of DTS and UJM using the setup described previously. All models here were trained with balanced interference mixtures ($p=1/2$). However, they were tested either under the same condition ($p=1/2$), or under a single-interference regime ($p=1$) similar to the setup in \cite{lancho2025rf}. We emphasize that even in the extreme case of $p=1$ (or $p=0$), uncertainty is still present, since $p$ is assumed to be unknown at inference time.

We begin with in-distribution BER performance, i.e., when training and testing are both conducted with balanced mixtures ($p=1/2$). When CS3 participates in the mixture (Fig.~\ref{fig:qpsk_cs3_mixtures}), DTS maintains an advantage over the 5L-UJM, where the UJM fully tracks its performance, emphasizing the importance of correctly scaling the UJM to converge to the DTS performance. However, for mixtures that are dominated by a difficult interference type mixtures, such as CS5G1 (Fig.~\ref{fig:qpsk_cs5g1_mixtures}), 5L-UJM suffices for both methods to have similar performance. These results indicate that UJM can replace multiple class-specific separators with a single DNN while preserving separation quality in diverse mixture scenarios.

Moving to the single-interference regime ($p=1$), Fig.~\ref{fig:qpsk+cs3} presents both BER and MSE vs. SINR performance, showing an easy-to-see example of how the MSE performance of the models ``transfers" to the BER performance. Moreover, UJM still tracks the performance of DTS, which confirms that the UJM has implicitly learned to specialize to the realized interference of the mixture, and not the interference mixture itself. Overall, these experiments highlight that UJM generalizes effectively under distribution shifts.\footnote{Additional results for the single-interference regimes with a QPSK SOI and interference mixtures reflect similar BER trends, and are omitted for brevity along with all MSE performances. These results are available at \cite{papergithub}.}
 %DTS, aided by the oracle detector, provides a strong reference since it deploys a dedicated separator per interference type.

% despite being trained on balanced mixtures.

% \begin{figure}[t]
%     \centering
%     \includegraphics[width=\linewidth]{figures/results/qpsk+emi1.pdf}
%     \caption{QPSK SOI with EMISignal1 interference only ($p=1$): BER and MSE vs. SINR for DTS and UJM.}
%     \label{fig:qpsk+emi1}\vspace{-0.4cm}
% \end{figure}

% \begin{figure}[t]
%     \centering
%     \includegraphics[width=\linewidth]{figures/results/qpsk+cs3.pdf}
%     \caption{QPSK SOI with CommSignal3 interference only ($p=1$): BER and MSE vs. SINR for DTS and UJM.}\vspace{-0.4cm}
%     \label{fig:qpsk+cs3}
% \end{figure}

% \begin{figure}[t]
%     \centering
%     \includegraphics[width=\linewidth]{figures/results/qpsk+emi1_or_cs2.pdf}
%     \caption{QPSK SOI with EMISignal1/CommSignal2 interference ($p=1/2$): BER and MSE vs. SINR for DTS and UJM.}
%     \label{fig:qpsk+emi1ORcs2}\vspace{-0.4cm}
% \end{figure}

\begin{figure*}
    \centering
    \includegraphics[width=0.925\linewidth]{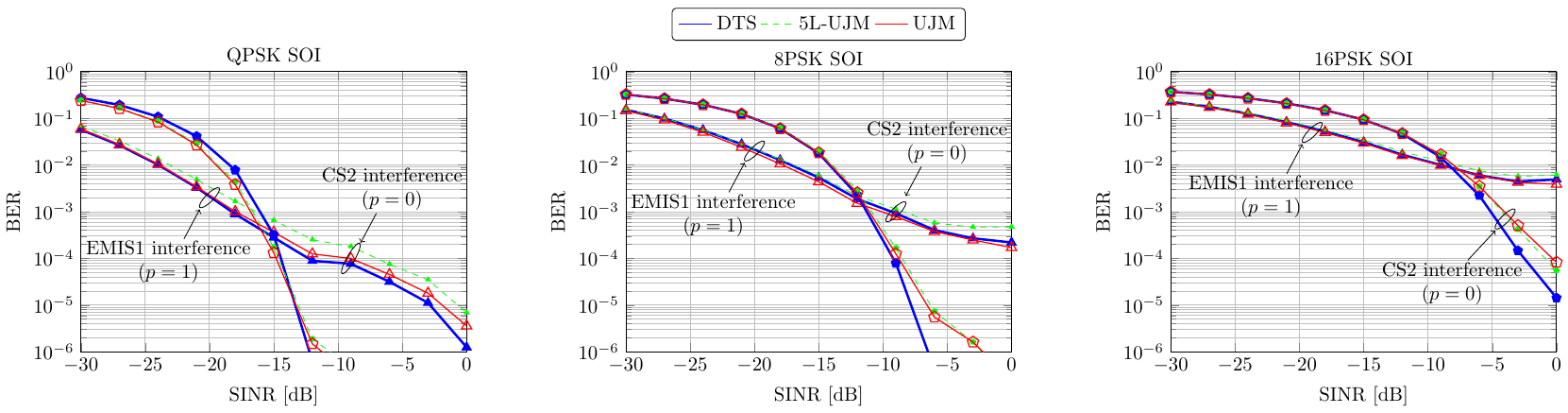}\vspace{-0.1cm}
    \caption{$M$PSK SOI with EMIS1 or CS2 interference ($p=1$ or $0$, respectively): BER vs. SINR for DTS, 5L-UJM and UJM.}\vspace{-0.6cm}
    \label{fig:mpsk_edgecase}
\end{figure*}

\begin{figure}
    \centering
    \includegraphics[width=0.925\linewidth]{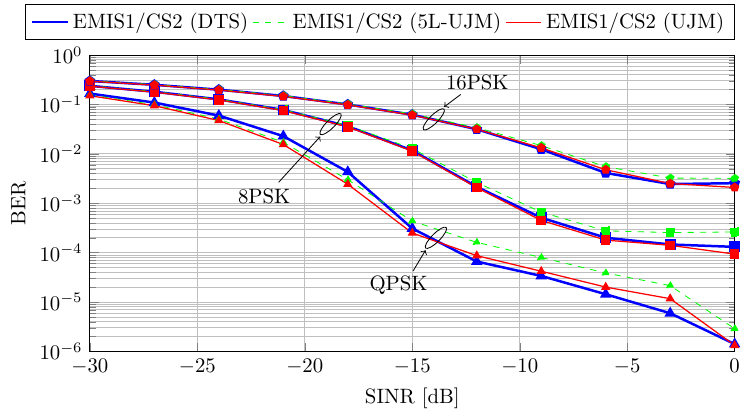}\vspace{-0.3cm}
    \caption{$M$PSK SOI with EMIS1/CS2 interference ($p=1/2$): BER vs. SINR for DTS, 5L-UJM and UJM.}
    \label{fig:mpsk_mix}\vspace{-0.6cm}
\end{figure}

To test the robustness of these conclusions, we extend the SOI to higher-order constellations (i.e., with higher information rates). Specifically, we compare QPSK, 8PSK and 16PSK with EMIS1/CS2 interferences. Figure \ref{fig:mpsk_edgecase} shows that, even in higher information rates, UJM generalizes well to the single-interference case ($p=1$ or $0$) despite having been trained on balanced mixtures ($p=1/2$). Figure \ref{fig:mpsk_mix} further shows results for the in-distribution case ($p=1/2$), where UJM matches DTS closely across most SINR levels. Both cases have small residual differences at high SINR that translate into negligible BER penalties. Overall, these experiments demonstrate that UJM is not limited to low-order modulations and can maintain effectiveness for richer statistical structures.

% \begin{figure}[t]
%     \centering
%     \includegraphics[width=\linewidth]{figures/results/16psk+emi1_or_cs2.pdf}
%     \caption{16PSK SOI with EMISignal1/CommSignal2 interference ($p=1/2$): BER and MSE vs. SINR for DTS and UJM.}
%     \label{fig:16psk+emi1ORcs2}\vspace{-0.4cm}
% \end{figure}

% \begin{figure}[t]
%     \centering
%     \includegraphics[width=\linewidth]{figures/results/16psk_marginal.pdf}
%     \caption{16PSK SOI with EMISignal1 or CommSignal2 interference ($p=1$ or $0$): BER and MSE vs. SINR for DTS and UJM.}
%     \label{fig:16psk_marginal}
% \end{figure}

% In summary, the numerical results show that UJM can effectively replace the DTS ensemble with a single deeper network, maintaining comparable performance across interference types, mixture conditions, and higher-order modulations. This validates the unified approach as a scalable alternative to detector-aided separation.

% \vspace{-0.25cm}
\section{Concluding Remarks}\label{sec:conclusions}
% \vspace{-0.15cm}
In this paper, we addressed the challenge of learning to separate RF signals under interference-type uncertainty. We formalized the problem as an extension of SCSS, and presented a decomposition of the MMSE estimator that highlights the role of implicit detection (via the posterior probabilities) of the interference type in optimal MMSE separation. Within a Gaussian mixture setting, we then established the asymptotic optimality of the DTS solution approach. Motivated by this analysis, we introduced the UJM, which avoids altogether type-specific detectors and specialized models by learning to jointly detect and separate within a single neural architecture, thus reducing labeling requirements and avoiding the linear scaling with the number of interference types that limits DTS.

Through simulations with multiple interference types and higher-order modulations, we showed that UJM can match the performance of an oracle-aided DTS ensemble across a range of conditions, including scenarios where the (mixture) distribution of the interference differs between training and testing. These findings establish UJM as a potential scalable and practical alternative to DTS, while raising new research questions, such as its behavior under large numbers of interference types, the role of architectural choices, and robustness beyond the low-SIR regime.

\appendices

\section{Lemma 1 Proof Outline}\label{app:lemma1}
Consider the suboptimal estimator
{\setlength{\belowdisplayskip}{4pt} \setlength{\belowdisplayshortskip}{4pt}
 \setlength{\abovedisplayskip}{4pt} \setlength{\abovedisplayshortskip}{4pt}
\begin{equation}
    \widehat{\rndk} \triangleq \arg \min_{k' \in \setT_K} \left|\psi_N(\rvecy,k')\right|.\label{eq:est-k}
\end{equation}}
\par\noindent Using the TDC, conditioned on the correct $\rndk=k$, we get that{\setlength{\belowdisplayskip}{4pt} \setlength{\belowdisplayshortskip}{4pt}
 \setlength{\abovedisplayskip}{4pt} \setlength{\abovedisplayshortskip}{4pt}
\begin{equation}
    \forall \ell \in \setT_K \backslash k:
 |\psi_N(\rvecy, \ell)| \xrightarrow[N\rightarrow\infty]{\text{a.s.}} c^2_{\ell}\neq 0,
\end{equation}}
% whereas conditioned on the correct $\rndk=k$, we can show that
\par\noindent where $c^2_{\ell}$ is the limiting value, whereas, we can show that{\setlength{\belowdisplayskip}{4pt} \setlength{\belowdisplayshortskip}{4pt}
 \setlength{\abovedisplayskip}{4pt} \setlength{\abovedisplayshortskip}{4pt}
\begin{equation}
    |\psi_N(\rvecy,k)| = \left|\tfrac{1}{N}\herm{\rvecy}\inv{\bC}_{\rndy_k} \rvecy-1\right| \xrightarrow[N\rightarrow\infty]{\text{a.s.}} 0.
\end{equation}}
\par\noindent Hence, as $N \to \infty$, the error probability of (\ref{eq:est-k}) is governed
by how far $|\psi_N(\rvecy,\rndk)|$ is from zero.

Formally, the error probability of \eqref{eq:est-k} is given by{\setlength{\belowdisplayskip}{4pt} \setlength{\belowdisplayshortskip}{4pt}
 \setlength{\abovedisplayskip}{4pt} \setlength{\abovedisplayshortskip}{4pt}
\begin{equation}
    \Prob\left(\widehat{\rndk} \ne \rndk\right) = \Prob\left(|\psi_N(\rvecy, \rndk)| > \underset{\ell \in \setT_K \backslash \rndk }{\min}|\psi_N(\rvecy,\ell)|\right).\label{eq:25}
\end{equation}}
\par\noindent Since $\widehat{\rndk}_{\text{\tiny MAP}}$ is optimal, we have $\Prob(\widehat{\rndk}_{\text{\tiny MAP}} \neq \rndk) \leq \Prob(\widehat{\rndk} \ne \rndk)$, hence we are left with the task of showing that \eqref{eq:25} is bounded away from zero and decreases at the desired rate, $o\left(N^{-\alpha}\right)$. The proof continues with a similar analysis as in \cite[App.~A]{proofs}.

\bibliographystyle{IEEEtran}

\end{document}